%% file: umap.tex
\documentclass[sigconf]{acmart}

\usepackage{booktabs} 
\usepackage{multirow} 
\usepackage{threeparttable} 
\usepackage{balance} 
\usepackage{stfloats}
\usepackage{amsmath}
\usepackage{wasysym}
\usepackage{soul}

\usepackage{pifont}
\newcommand{\cmark}{\ding{51}}%
\newcommand{\xmark}{\ding{55}}%

\setcopyright{rightsretained}
\newif\ifworkinprogress
\workinprogressfalse

\ifworkinprogress
  \newcommand{\rev}[1]{\textcolor{red}{ #1}}
  \newcommand{\revi}[1]{\textcolor{blue}{ #1}}
\else
  \newcommand{\rev}[1]{#1}
  \newcommand{\revi}[1]{#1}
\fi

\setcopyright{none}
\renewcommand\footnotetextcopyrightpermission[1]{} 
\begin{document}
\settopmatter{printacmref=false}
\title{\revi{Revisiting Interest Indicators Derived from Web Reading Behavior for Implicit User Modeling}}


\author{Mirjam Augstein}
\affiliation{%
  \institution{University of Applied Sciences Upper Austria \\
  Communication and Knowledge Media}
  \streetaddress{Softwarepark 11}
  \city{Hagenberg}
  \country{Austria}
  \postcode{4323}
}
\email{mirjam.augstein@fh-hagenberg.at}

\author{Johannes Sch\"onb\"ock}
\affiliation{%
  \institution{University of Applied Sciences Upper Austria\\
  Department of Communication and Knowledge Media}
  \streetaddress{Softwarepark 11}
  \city{Hagenberg}
  \country{Austria}
  \postcode{4323}
}
\email{johannes.schoenboeck@fh-hagenberg.at}

\author{Christina Lettner}
\affiliation{%
 \institution{University of Applied Sciences Upper Austria \\ Research and Development}
 \streetaddress{Softwarepark 11}
 \city{Hagenberg}
  \country{Austria}
  }

\author{Josef Altmann}
\affiliation{%
 \institution{University of Applied Sciences Upper Austria \\ Communication and Knowledge Media}
 \streetaddress{Softwarepark 11}
 \city{Hagenberg}
 \country{Austria}
  }
\email{josef.altmann@fh-hagenberg.at}

\renewcommand{\shortauthors}{M.~Augstein et al.}

\begin{abstract}
\revi{Today, intelligent user interfaces on the web often come in form of recommendation services tailoring content to individual users.} Recommendation of web content such as news articles often requires a certain amount of explicit ratings to allow for satisfactory results, i.e., the selection of content actually relevant for the user. Yet, the collection of such explicit ratings is time-consuming and dependent on users' willingness to provide the required information on a regular basis. Thus, using \emph{implicit interest indicators} can be a helpful complementation to relying on explicitly entered information only. Analysis of reading behavior on the web can be the basis for the derivation of such implicit indicators. \rev{Previous work has already identified several indicators and discussed how they can be used as a basis for user models.} \revi{However, most earlier work is either of conceptual nature and does not involve studies to prove the suggested concepts or relies on meanwhile potentially outdated technology. All earlier discussions of the topic further have in common that they do not yet consider mobile contexts.}
\revi{This paper builds upon earlier work, however providing a major update regarding technology and web reading context, distinguishing between desktop and mobile settings. This update also allowed us to identify a set of new indicators that so far have not yet been discussed. This paper describes (i) our technical work, a framework for analyzing user interactions with the browser relying on latest web technologies, (ii) the implicit interest indicators we either revisited or newly identified, and (iii) the results of an online study on web reading behavior as a basis for derivation of interest we conducted with 96 participants.} 
\end{abstract}

%
%



\maketitle


\input{introduction}

\input{relatedwork}

\input{indicators}

\input{implementation}
\input{userstudy}

\input{usermodels}
\input{impact}
\input{conclusions}

\bibliographystyle{ACM-Reference-Format}
\balance
\bibliography{literature}

\end{document}

%% file: introduction.tex
\section{Introduction}
\label{sec::introduction}
As in the previous years the web was constantly growing, it gets more and more important to understand users to be able to present relevant information to them. \revi{In order to implement more intelligent websites,} different kinds of interest indicators can be helpful when predicting user interest in certain web content elements, e.g., news articles.
In general, user interests can be gathered in (i) an \emph{explicit} manner \cite{jawaheer2010}, i.e., asking questions to users, (ii) an \emph{implicit} manner \cite{Hijikata2004}, i.e., information is gathered by a user's behavior or in (iii) a \emph{hybrid} manner, i.e., a combination of implicit and explicit means. Both, explicit and implicit ratings or indicators can provide a good foundation for recommendation of other, previously unseen web content elements. However, implicit means relieve the user of the burden of rating every content on its interestingness \cite{Zahoor2014}.  
Concerning gathering user interests on the web, one may further distinguish between (i) \emph{server-side} and (ii) \emph{client-side} profiling \cite{Bilenko2011}. While server-side approaches basically rely on analyzing HTTP requests, client-side approaches allow for continuous monitoring and a more profound analysis of user interaction \cite{Hauger2008}. 

Therefore, this paper focuses on a \emph{client-side} analysis of \emph{implicit interest indicators} derived from web reading behavior. 
It builds upon earlier work (see e.g., \cite{Claypool2001,Hauger2011,Zahoor2014,Zahoor2015}), \revi{revisiting known indicators with state of the art web technologies. During this process, we could additionally identify novel indicators that have not been discussed so far. Further, our work contributes new insights as it explicitly distinguishes between desktop and mobile contexts which has not been the case in earlier research.}

The paper further presents the results of a study conducted with 96 web users. \revi{The study aimed at determining} the informativeness and suitability of particular implicit interest indicators (i) for web users in general and (ii) for individual web users in particular. 
We provide insights into the indicators themselves (e.g., their potential to reliably predict user interest), how they may be obtained by JavaScript (JS) and additionally suggest a \revi{possible base} structure for user models relying on reading behavior. \revi{This structure can be considered as a foundation for subsequent work}. Further, \rev{we envision} the results of our study as \rev{a basis for} later recommendation of content \revi{in intelligent web user interfaces}. 



%% file: relatedwork.tex
\section{Related Work}
\label{sec:relatedwork}
This section describes related work on (i) technologies underlying client-side analysis of implicit interest indicators, (ii) the indicators themselves, and (iii) user studies related to implicit interest analysis. 

\textbf{Technologies.} Early approaches were based on custom browsers \cite{Claypool2001,Stephanidis1997} or browser plugins \cite{Goecks2000}. Whereas these approaches do not suffer from any limitations of web browsers, the application thereof is hindered since the user needs to install additional software. In contrast, more recent approaches base on JS and AJaX, which is commonly supported in today's web browsers and allows sending data to a web server for persisting it in a user profile. 
\rev{A vision of such systems is presented in  e.g.,~\cite{Hauger2008, Hauger2009b, Zahoor2015}}.
However, none of the earlier approaches explicitly mentioned the inclusion of mobile browsers on tablets or smart-phones although nowadays more and more users use the ``mobile web''. Thus, our approach puts special emphasis on including mobile browsers and how to obtain interest indicators during surfing on mobile devices (cf.~Section~\ref{sec:indicators} and \ref{sec:implementation}).

\textbf{Implicit interest indicators. } 
\rev{A first high level categorization of implicit interest indicators distinguishing between \textit{marking} (e.g., bookmarking), \textit{manipulation} (e.g., cutting and pasting or searching), \textit{navigation} (e.g., following a link), \textit{external} (e.g., a user's heart rate or temperature) and \textit{repetition indicators} (e.g., spending \textit{more} time on a page or doing \textit{lots} of scrolling) is presented in ~\cite{Claypool2001}}. 

\rev{First approaches to derive implicit indicators focused on the whole web page as a single source of information resulting in rather coarse-grained indicators}.
\rev{In \cite{Hauger2009}}, fragmentation mechanisms (e.g., \textit{split by vertical position}, \textit{split by content type} or \textit{split by structural information}) \rev{were proposed} to allow for a more fine-grained analysis. This led to new interest indicators as well, i.e., \textit{visible time}, \textit{mouse over time}, \textit{mouse on same y time}, \textit{number of mouse moves}, \textit{number of clicks} and \textit{number of text selections}.
Based on these considerations, 
an algorithm to predict which paragraphs in a HTML document have been read, \rev{was presented \cite{Hauger2011}}. 

\rev{In \cite{Smucker2014}, special emphasis was put on indicators derived from mouse movements.} 
\rev{There,} a classification scheme for mouse movements distinguishing between \textit{decision-only} (i.e., mouse movement related to an explicit rating such as moving the mouse to a ``relevant'' button), \textit{horizontal}, \textit{vertical}, \textit{highlighting}, \textit{re-scoping} (i.e., a movement related to changing the mouse position after loading a new page), \textit{scrolling}, \textit{random}, and \textit{no-movement} \rev{was proposed}.

\rev{In \cite{Zahoor2014}}, another categorization of implicit interest indicators that, on the top level, 
involves \textit{time} (active and passive), \textit{mouse} (clicking, scrolling and movements) and \textit{keyboard} (movements via cursor keys and shortcuts) \rev{was discussed}. This categorization \rev{builds} the \rev{primary} basis for our approach described in this paper. The same authors stated in \cite{Zahoor2015} that a relative ordering and a combination of implicit indicators may further improve relevancy.

Although a lot of valuable research has been conducted on implicit interest indicators, to the best of our knowledge, mobile devices and indicators thereon have not been considered up to now. Consequently, in this paper we discuss to which respect indicators analyzed for desktop usage may be applied also on mobile devices and which new indicators result from mobile devices. 

\textbf{User Studies.}
A first user study with 72 students on analyzing the correlation between implicit and explicit interest indications was conducted \rev{in}~\cite{Claypool2001}. Yet, this study merely focused on time-based indicators (i.e., \textit{time spent on a page}, \textit{time spent moving the mouse}, \textit{number of mouse clicks}, and \textit{time spent scrolling}). The results suggest that \textit{time spent on a page} and \textit{time spent scrolling} are good indicators of interest. Regarding \textit{time spent moving the mouse}, a positive relationship between the indicator and the explicit rating could be found but it is expressive mainly for determining which pages have the least amount of interest. For distinguishing between higher interest levels, the indicator is not accurate. For \textit{number of mouse clicks}, no correlation with the explicit rating could be found.
 
 Reading behavior as basis for the derivation of user interest was further analyzed 
 \rev{in}~\cite{Hauger2008,Hauger2009,Hauger2009b,Hauger2011}. In \cite{Hauger2009}, a study with 53 users comparing the results of client-side user monitoring and explicit user feedback is presented. A single page containing news items was used for the study and the following indicators related to mouse activity were used to determine whether an item has been read: \textit{mouse time above item}, \textit{mouse time on same y as item}, \textit{time item is visible}, \textit{amount of mouse moves above item}, \textit{number of clicks on item}, and \textit{number of text selections in item}. The results suggest that generally, the observation of client-side activity at least doubles the probability that an item has been read (80\% of the items with no activity had not been read). Yet, 
 \rev{the authors} also found that about half of the items with interaction times or mouse movements have been skipped as well \cite{Hauger2009}. Regarding the individual indicators, the study revealed a positive correlation between \textit{mouse over time} and a higher probability the item has been read. Further, \textit{mouse moves}, \textit{click events} and \textit{text selection} have been shown to be strong indicators while \textit{visibility time} and \textit{vertical mouse position} did not allow for sufficient conclusions. The study design \rev{in \cite{Hauger2009}} is similar to ours, however, we did not mainly aim at predicting whether an item has been read or not, but how interesting it was (as a basis for future recommendation). Further, we integrated more indicators (including such relevant for a mobile context). 
 
 In \cite{Hauger2011}, an eyetracking study \rev{was conducted in order} to determine whether content elements have been read or not. Eyetracking in this case provides a more reliable explicit source of information, compared to users' statements. The study e.g., revealed that prediction of a user's gaze position is more accurate while the mouse is in motion and that mouse actions (such as text selections, clicking or mouse movements) allow for better predictions but these activities occur only for a fraction of the total observation time. Related to this, in our study, we observed that user behavior related to mouse activity is highly individual (thus, the related indicators perform well for some users but fail for others).
 
 
More recently, a study with 50 participants, analyzing implicit interest indicators based on \textit{mouse} (i.e., \textit{movement}, \textit{click} and \textit{scroll}) and \textit{keyboard} (i.e.,\textit{shortcuts} and \textit{movements}) activities \rev{was conducted \cite{Zahoor2015}}. The indicators were analyzed \rev{by collecting} the users' explicit statements \rev{since the users were asked} to rate their behavior regarding mouse or keyboard activities. E.g., users had to rank different mouse actions (movement, click or scroll) based on their frequency of usage. This approach might be error-prone as it is dependent on users' realistic answers. We thus rely on automated analysis of user activities on a web page only.


\rev{In \cite{Smucker2014}}, an analysis on data of an earlier study \cite{smucker2010} with 48 participants that had to judge the relevance of news documents and document summaries \rev{was conducted}. Additionally, different kinds of mouse movements (\textit{decision only}, \textit{horizontal}, \textit{vertical}, \textit{highlighting}, \textit{re-scoping}, \textit{scrolling}, \textit{random} and \textit{no-movement}) were captured as implicit indicators. While the classification of mouse movements is interesting, we believe the results might be biased by the way how users had to rate the relevance of the documents, i.e., through clicking a ``relevant'' or ``not relevant'' button on top of the page. Most users thus moved only between these buttons which is why more general conclusions based on mouse activity inside the actual text area are not possible. We thus separated the reading area from the area to provide explicit feedback in our study.


%% file: indicators.tex
\section{Implicit Interest Indicators}
\label{sec:indicators}
In the following, we present implicit interest indicators that we considered in our implementation (cf.~Section~\ref{sec:implementation}) as well as in our user study (cf.~Section \ref{sec:userstudy}), mainly based on the indicators suggested in \cite{Hauger2011,Zahoor2015}. We categorize the indicators by their source of origin as well as their support on desktop or touch-based mobile devices (cf. Table~\ref{tab:overviewIndicators}). After describing the general intention, we briefly outline how the indicators may be detected by means of \rev{recent} JS \rev{technology (including plain JS as well as frameworks)}.


\begin{table}[!b]
  \caption{Overview of implicit indicators}
  \label{tab:overviewIndicators}
  \begin{tabular}{cccc}
    \toprule
    & \textbf{Indicator} & \textbf{Desktop} & \textbf{Mobile} \\
    \midrule
   \textbf{Time} & Visibility & \cmark & \cmark\\
   \midrule
   \multirow{6}{*}{\textbf{Movement}} & Random Movement & \cmark & \xmark \\
   & Move in Fragment & \cmark & \xmark\\
   & Mouse over Fragment & \cmark & \xmark\\
   & Horizontal Movement & \cmark & \xmark\\
   & Vertical Movement & \cmark & \xmark\\
   & Mouse on Same Y & \cmark & \xmark\\
  \midrule
  \textbf{Contact} & Contact in Fragment & \cmark & \cmark \\
  & {Contact on Same Y} & \xmark & \cmark\\
   \midrule
   \multirow{2}{*}{\textbf{Text}} & Select & \cmark & \cmark \\
   & Cut \& Copy & \cmark & \cmark\\ 
   \midrule
   \multirow{3}{*}{\textbf{Other}} & Zoom & \cmark & \cmark \\
   & Swipe & \xmark & \cmark\\ 
   & Orientation Change & \xmark & \cmark\\ 
  \bottomrule
\end{tabular}
\end{table}

\subsection{Time-Based Indicators}
\label{sec:timeIndicator}

\textit{Time-based indicators} try to derive user interests by means of the time spent on a website or visibility time of a fragment. By a fragment we mean a specific part of a website, e.g., a paragraph or a div-container, which can be defined by a programmer by assigning an explicit class (cf. Section~\ref{sec:implementation} for details). In general it has been proven that the more time a user spends on a website the more interesting it is to him/her \cite{Claypool2001,Faucher2011,Kellar2004,Rastegari2010,Velayathan2007,Zahoor2014}. The time spent on a website may be split into \emph{active} and \emph{passive} intervals. Active interval means the duration a user is reading a text or interacting with the website whereas passive interval means the time where the website is shown in the browser but there is no user interaction (it is assumed that the user is distracted from the website during this period of time). 
In JS, the elapsed time between two events is compared. If it exceeds a certain delta (60 sec.~by default) it is considered as a passive, otherwise as an active interval. 


\textbf{Visibility.} This indicator determines how often and how long (cf.~\textit{Visibility Count} and \textit{Visibility Seconds} in Section~\ref{sec:results}) a fragment of a website has been visible, \rev{i.e., it has been within the viewport of a user}. \revi{This is especially relevant for handling vertical and horizontal scrolling irrespective of the source of origin, i.e., it does not matter if scrolling is indicated by the mouse wheel, a page down button press, by using the scrollbar or by swiping on mobile phones. Furthermore, we do not solely count the number of scroll actions as done e.g., in \cite{Claypool2001}, but focus on the actual visibility time of a fragment in order to derive if the text has been read or not depending on the length and difficulty of a text within a fragment \cite{Hauger2009}}.

Every time the visible part of a website changes, the \texttt{scroll} event is thrown in JS, both on desktop and mobile devices. Within the according event handler for every fragment of a web page its size and its relative position to the viewport are calculated which is than compared with the current viewport of the browser window to calculate the active visibility time of a certain fragment using the timestamps of the scroll event. The combination of time spent and scrolling has been shown as a valuable indicator in \cite{Kvrivz2012}.

\subsection{Movement-Based Indicators}
\label{sec:moveIndicator}
Indicators that stem from mouse movements are subsumed under \textit{movement-based indicators}. Since there is no mouse pointer on mobile devices, these indicators apply to desktop devices only. 

\textbf{Random Movement and Move in Fragment.} It is assumed that the more frequent mouse movements within a fragment are, the more likely the user is also looking at a certain fragment \cite{Mueller2001,Torres2008}. Yet, mouse movements might not be a very good indicator that a certain fragment is of interest, but no movements strongly indicate that it has been skipped and is thus not interesting to the user \cite{Smucker2014}.

\textbf{Mouse over Fragment.} Studies revealed that users tend to place the mouse within a text they are currently reading \cite{Hauger2011,Smucker2014}. This indicator measures how often and how long (cf.~\textit{Mouse over Fragment Count} and \textit{Mouse over Fragment Seconds} in Section~\ref{sec:results}) the mouse is placed over a certain fragment to derive the interest depending on the length and difficulty of the text.

\textbf{Horizontal Movement.} Horizontal movements of a mouse, if happening regularly and over several lines of a text, might indicate that a user uses the mouse as a ``pointer'' for reading \cite{Smucker2014}.

\textbf{Vertical Movement.} Similar to horizontal movements, vertical movements of a mouse might indicate that users use the mouse to ``mark'' the current line of text that they are reading \cite{Smucker2014}. 

\textbf{Mouse on Same Y.} Whereas the last two indicators assume that a user is moving the mouse along the text when reading, this indicator relates to users that place the mouse e.g., on the right side of a website, when starting to read but do not move the mouse during reading. Those fragments that are on the same horizontal line, i.e., y-coordinate, are assumed to be of interest \cite{Hauger2009}. 

\textbf{Detection in JS.} To detect \textit{movement-based indicators} the mouse events of JS need to be handled. In particular, the \texttt{mousemove} event can be used to track all movements of the mouse. For \textit{fragment-based indicators} the \texttt{mouseenter} and \texttt{mouseleave} events may be used in order to count how often and how long a fragment has been visited. Horizontal and vertical movements may be detected by comparing the \texttt{pageX} and \texttt{pageY} attributes of the mouse events. However, since users are not able to move the mouse in a straight horizontal/vertical direction some fuzziness needs to be respected. For the indicator \textit{Mouse on Same Y}  additionally the \texttt{scroll} event is considered to check for new fragments after scrolling.

\subsection{Contact-Based Indicators}
\label{sec:clickIndicator}
These indicators relate to click (desktop) and touch (mobile) activity. \rev{The latter is a novel addition to earlier collections of implicit interest indicators and has not been analyzed before}.

\textbf{Contact in Fragment.} This indicator recognizes a click (irrespective of right/left mouse clicks on desktop or touch/press on mobile devices) in a certain fragment. Especially if a link is clicked one may conclude that a fragment is of interest to a user \cite{Faucher2011,Shen2005, Velayathan2007}. 

\textbf{Contact on Same Y.} Similar to placing a mouse near the text of interest, we assume that a user might use a touch interaction to ``mark'' the fragments on the same horizontal line for reading.

\textbf{Detection in JS.} A click can be recognized by attaching \texttt{click} event handlers (desktop devices). For mobile devices (\texttt{tap} and \texttt{press} (long tap) handlers) we utilized the external library hammer.js\footnote{https://hammerjs.github.io/}. 

\subsection{Text-Based Indicators}
\label{sec:textIndicator}

\textbf{Select, Cut and Copy.} Although the selection and copying of a text occurs relatively seldom compared to the other indicators, studies found out that a selected text indicates high relevancy \cite{Hauger2009,Hijikata2004,Zahoor2014}. The \texttt{mouseup} event on desktop and the \texttt{touchend} event on mobile devices may be caught. \rev{Additionally}, it needs to be checked if the according event object contains a so called \texttt{selection} object which contains the selected text. Furthermore, on desktop also the \texttt{keyup} event needs to be registered in order to check for text selection via the keyboard (e.g., by Shift + cursor keys). If a text is cut\footnote{On a website it is not possible to actually ``cut'' a text, thus the event is treated as equal to copy.} or copied, JS provides the \texttt{cut} and \texttt{copy} events which also contain this text. 

\subsection{Other Indicators}
\label{sec:otherIndicator}
Here we describe further indicators rather specific for the mobile context. \rev{All of them are novel contributions to work on implicit interest indicators and have not been analyzed in this context before}.

\textbf{Zoom.} If a user is interested in a certain fragment, he/she might try to enlarge this fragment. Although zooming is not commonly used on desktop, we assume it is especially important on mobile devices. To detect a zoom on mobile devices, we again utilize hammer.js which provides a \texttt{pinch} event. Besides the information from the event itself, it is also necessary to detect the fragments that are visible after zooming, which is done as described for \textit{visibility}. 

\textbf{Swipe.} Swipe means a fast scrolling on mobile devices. We assume that fragments that have been skipped during swiping might be of no interest since the user can not see them. This is why we explicitly considered swiping in contrast to scrolling which allows at least some skimming of a text. Still, scrolling is part of several other indicators.
To detect this indicator we use jQuery Mobile\footnote{https://jquerymobile.com/} which provides \texttt{scrollstart/scrollend} as well as a \texttt{swipe} event. When starting (\texttt{scrollstart}) or stopping (\texttt{scrollend}) a swipe, the visible fragments are temporarily saved in two separate lists. 
All fragments during a swipe are saved in an extra list which is then used to build the intersection with the lists of visible fragments at the beginning/ending of the swipe to obtain the skipped fragments.

\textbf{Orientation Change.} Mobile users sometimes turn their mobile phones from portrait to landscape view if they are reading a certain fragment and back to portrait if they are finished. Again, in combination with others, this indicator might increase certainty that a fragment is of interest to a user. Modern browsers support the JS event \texttt{orientationchange}, otherwise jQuery Mobile provides a workaround. Besides the data from the event, it is also necessary to check for the visible fragments after the orientation changed.

%% file: implementation.tex
\section{Implementation}
\label{sec:implementation}

\begin{figure}[!htbp]
\centering
\includegraphics[width=.95\linewidth,keepaspectratio]{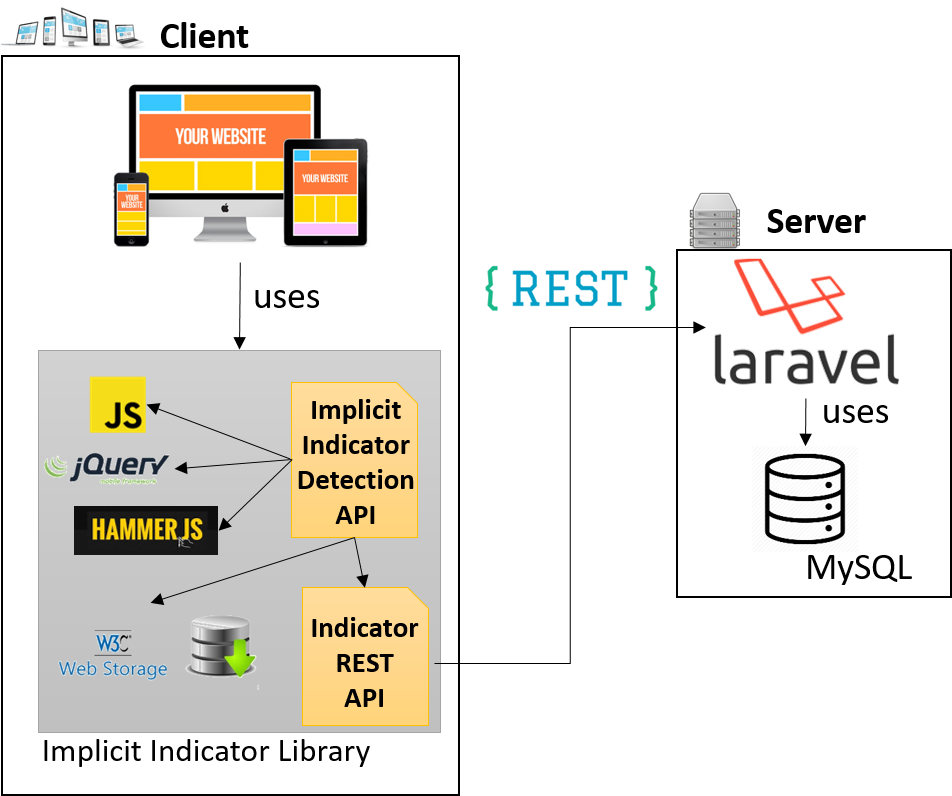}
\caption{Overview on Prototype Architecture}
\label{fig:architecutre}
\end{figure}

This section presents the general architecture of our current prototype (cf.~Figure~\ref{fig:architecutre}) \rev{as another contribution of the paper}.
The prototype may roughly be divided into a client-side \emph{Implicit Indicator Library} \rev{component} and a \textit{REST API} as well as the \textit{database} on the server side as described in the following.

\subsection{Implicit Indicator Library}
\label{sec:JSLib}
The core component of our prototype is a JS library to obtain the indicators described above (cf.~\textit{Implicit Indicator Detection API} in Figure~\ref{fig:architecutre}). To apply the library to an HTML document -- in our user study a Typo3-based news site -- the programmer needs to assign a unique CSS class (configurable via the library settings) to the HTML elements that should be handled as a fragment. During initialization, the fragments of the HTML documents are automatically obtained, whereby we respect the hierarchical ordering of the DOM tree, e.g., if an indicator is detected on a child fragment it is also propagated to its parent fragments. However, if solely the actual fragment where the indicator occurred should be tracked, this may again be configured in the library settings. Further, the library may be configured based on which indicators are of interest for the programmer, i.e., it is possible to enable/disable tracking of indicators. To minimize REST calls to the server (cf.~Section~\ref{sec:server}) and to omit a loss of data in case that internet connection is lost 
or the server is offline, the Web Storage API\footnote{https://developer.mozilla.org/de/docs/Web/API/Web\_Storage\_API} is used to provide an intermediate data store on the client. Thus, not every event or implicit indicator is immediately pushed to the server but only after a certain time span (30 sec.~by default) or the maximum number of local events/indicators occurred (50 by default).

\subsection{REST API and Database}
\label{sec:server}
\begin{figure}[!ht]
\centering
\includegraphics[width=1.0\linewidth,keepaspectratio]{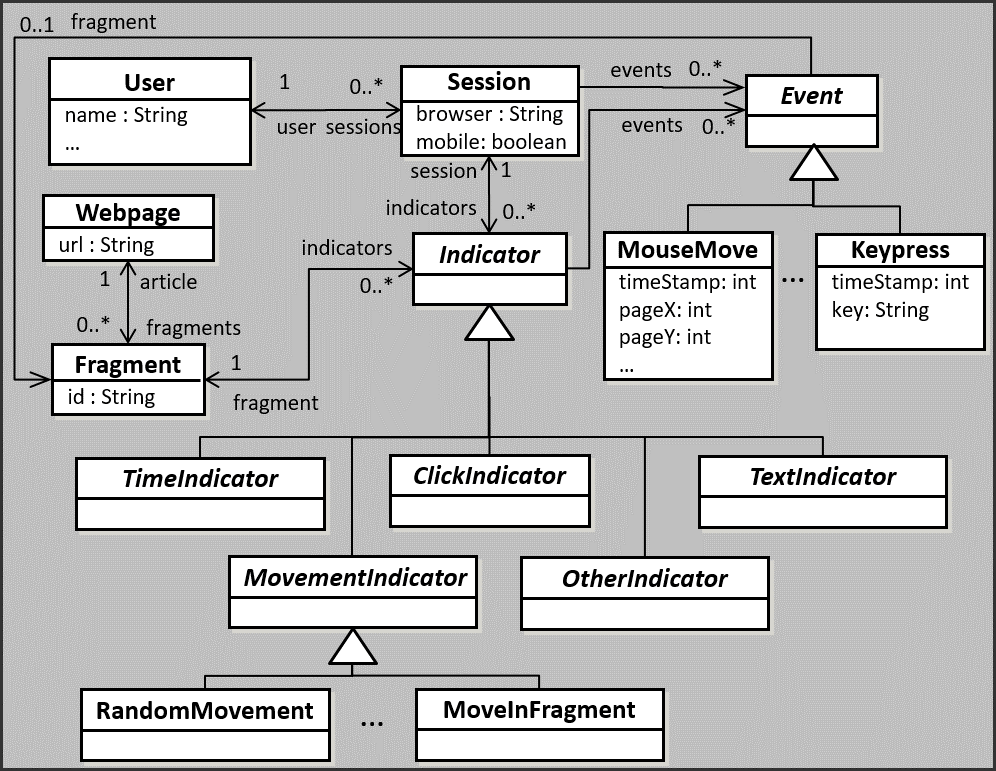}
\caption{Extract of Database as UML Diagram}
\label{fig:db}
\end{figure}

To persist the data of the detected events and indicators in a database on a server, we make use of the REST protocol. On the client side, we provide functions which wrap the data of the events/indicators into according JSON objects which are then used by respective POST methods to submit the data to the server (cf.~\textit{Indicator REST API} in Figure~\ref{fig:architecutre}). On the server side, the according REST API is defined using the Laravel\footnote{https://laravel.com/} framework. Besides methods that store the events and indicators in the underlying MySQL database also means for authentication are provided to assign the occurring events or indicators to users.
Figure~\ref{fig:db} shows an extract of the database in terms of an UML diagram. A \texttt{User} is connected to \texttt{Session}s, which represent a certain interaction with a website via a specific browser. Furthermore, the actual \texttt{Webpage} and its \texttt{Fragment}s are stored in the database. In case a user interacts with a web page, the emerging \texttt{Events} and/or \texttt{Indicators} are stored in the database and linked to the according \texttt{Fragment} instance. For \texttt{Indicator}s as well as \texttt{Event}s according sub-classes exist. However, for reasons of brevity we omitted all the concrete sub-classes and simplified the relations between \texttt{Indicator}s and \texttt{Event}s (e.g., to a \texttt{MoveInFragement} indicator only \texttt{MouseMove} events may be assigned).

%% file: userstudy.tex
\section{User Study}
\label{sec:userstudy}
Here we describe the user study conducted to evaluate the indicators (cf. Section~\ref{sec:indicators}) and their \rev{potential} for predicting user interest in web content.

\subsection{Setting}
The study was online for three weeks (end of September to mid of October 2018). 
We set up a responsive Typo3-based news page to track the indicators with the JS library presented in Section \ref{sec:implementation}.
The website was fully text-based to prevent additional effects stemming from the nature of pictures or other visual kind of information presentation. 
It contained a mix of teaser texts linked to the full-text page of 20 news articles related to different topics (sports, health, lifestyle, technology, and travelling). 
The articles were all similar regarding length and readability. For the latter, we used Flesch's reading ease test \cite{flesch48} and selected articles with similar readability scores. On a scale between 0 and 100, values between 100 and 70 are considered \textit{easy} (in three gradations: very easy, easy and fairly easy), values between 69 and 60 are considered \textit{standard} and values between 59 and 0 are considered \textit{difficult} (again in three gradations: fairly difficult, difficult and very confusing). For the selection of our news articles we avoided texts that can be considered \textit{very confusing}. All selected texts had readability scores between $80.4$ and $42.0$ ($M=63.77$, $SD=10.92$). The teaser texts had readability scores between $75.3$ and $46.1$ ($M=60.46$, $SD=8.33$). Of the 20 articles, 17 were in a ``moderate'' area with values between 79 and 50. This way we aimed at preventing a significant bias due to text complexity. Further, we selected articles that are nearly timeless (so they would not date out during the study's online time). 

\subsection{Procedure and Methodology}
\label{sec:procedure}
After general instructions and information on the study (e.g., related to privacy issues and data processing) the participants had to scan the overview page and take as much time as they needed to read as many articles as they liked in further detail as we wanted to be able to record as-natural-as-possible reading behavior. Since we automatically derived information about the device from the browser used, participants could freely chose the device (desktop or a mobile). 
 After the participants finished reading, they answered a web-based questionnaire which we set up using the Unipark\footnote{https://www.unipark.com/} survey tool. The questionnaire contained (i) questions related to demographic information (age, gender), (ii) questions regarding general news reading behavior, (iii) information related to usage of devices like smart phones and desktops, and (iv) questions related to the articles themselves (whether they had noticed the individual articles and if yes, how interesting they found them on a seven-point Likert scale). Participation took about 20 minutes on average (about 15 minutes for reading and about five for the survey). 

Before the actual user study, we conducted a thorough multi-phase pretest with five selected users that did not participate in the study later. \revi{The aims of the pretest were twofold. First, the pretest allowed us to observe different kinds of user behavior during browsing our news page and reading. It helped us to identify types of actions that might happen and what meaning these actions might have for the users. For instance, one of our pretest users only interacted with the web page when necessary (e.g., to scroll down after the bottom of the page had been reached) while another user continuously scrolled the text to keep it in approximately the middle of the screen. Other users often moved the mouse to the position of the text they were currently reading. Second, the pretest helped us to gain first insights about the indicators themselves. Third, we were able to (i) test the technical infrastructure in a study setting, (ii) ensure users understand instructions, tasks and questions without further assistance, and (iii) gain feedback related to the design of the website used for the study}.
During the first phase, two users underwent the full study procedure on a desktop device including a Tobii T40 Eyetracker.
After this phase we changed (i) page layout, (ii) login process\footnote{Users needed to register (solely username, no password) in order to link the reading behavior with the externally hosted questionnaire.}, (iii) design of the questionnaire, and (iv) in one case, selection of news articles. The second phase included (i) another test with an additional user on the eyetracking device and (ii) two tests with a user of the first phase to validate the introduced changes. 
After finalizing the desktop version of the study we did a pretest on smart phones with two more users resulting in a revision of the questionnaire's layout for mobile devices. 

\begin{table}[!b]
  \caption{Correlation (Pearson's \textit{r}) overview for all indicators in the desktop setting.  * or ** denote statistical significance (on the .05 or .001 level).}
  \label{tab:resultsoverviewdesktop}
  \begin{tabular}{p{3.2cm}p{0.92cm}p{0.92cm}p{0.92cm}p{0.92cm}}
    \toprule
    \textbf{Indicator} & \textbf{r}\newline \tiny{Overview Page} & \textbf{p} \newline \tiny{Overview Page} &  \textbf{r} \newline \tiny{Detail Page} & \textbf{p} \newline \tiny{Detail Page}\\
    \midrule
    \textbf{Visibility} & & & \\
    Count & .032& .179 & .373 & < .001**\\
    Seconds & .086 & < .05* & .116 & < .001**\\
    \midrule
    \textbf{Random Movement}  & .173 & < .001** & .199 & < .001**\\
    \midrule
    \textbf{Move in Fragment}  & .172 & < .001** & .120 & < .001**\\
     \midrule
    \textbf{Mouse over Fragment} &  &  & & \\
    Count & .180 & < .001** & .088 & < .05*\\
    Seconds  & .203 & < .001** & .220 & < .001**\\
    \midrule
     \textbf{Horizontal Movement} & .131 & < .001** & .090 & < .05*\\
    \midrule
    \textbf{Vertical Movement} & .092 & < .05* & .198 & < .001**\\
    \midrule
    \textbf{Mouse on Same Y}  &  &  &  & \\
    Count  & .023 & .251 & .061 & < .05*\\
    Seconds  & .294 & < .001** & .121 & < .001**\\
    \midrule
    \textbf{Contact in Fragment}  & .208 & < .001** & .150 & < .001**\\
  \bottomrule
\end{tabular}
\end{table}
 
\subsection{Participants}
\rev{We recruited} 109 users via mailing lists of the authors' university as well as mailing lists of the \revi{user modeling} research community. \rev{These users} had activity on the news page and logged in to the questionnaire. 
Of these, 96  
finished the questionnaire (the remaining 13 had to be excluded \rev{due to missing data}). Participation was fully anonymous, users who wanted to have a chance of winning one of 20 Amazon vouchers, voluntarily provided their email address \rev{at the end of the study (which was not linked to the rest of their data)}.  
\rev{Of all 96 participants, 35 were female, 61 male}. 74 participants (23 female, 51 male) used a desktop device, 22 (12 female, 10 male) a smart phone.  
They were between 18 and 65 years old ($M=28.85$, $SD=10.11$). Split up into participants with desktop and mobile devices, desktop participants were between 19 and 65 years old ($M=29.82$, $SD=10.86$), mobile participants were between 18 and 38 years old ($M=25.59$, $SD=6.13$). 
Asked for their laptop/PC usage habits, 68 participants ($70.8\%$) use such devices more than five hours a day, eight ($8.3\%$) between four and five hours, six ($6.3\%$) between three and four hours, eight ($8.3\%$) between two and three hours, three ($3.1\%$) between one and two hours and the remaining three ($3.1\%$) less than one hour a day.
Further, eleven of 76 participants ($11.5\%$) use a smart phone more than five hours a day, eleven ($11.5\%$) use it between four and five hours, 20 ($20.8\%$) between three and four hours, 20 ($20.8\%$) between two and three hours, 23 ($24\%$) between one and two hours and eleven ($11.5\%$) less than one hour.
Concerning news reading behavior, eleven participants ($11.5\%$) stated to read news several times a day, seven ($7.3\%$) daily, 22 ($22.9\%$) several times a week, 27 ($28.1\%$) once a week, and 29 ($30.2\%$) rarely or never.

\begin{table*}[!t]
  \caption{Statistically significant correlations of interest indicators with explicit user interest assessment. All values except mean correlation (values between -1 and 1) are in absolute numbers (out of 74 desktop users), measured for the news overview page and detail page. Not statistically significant correlations are not included. 
  }
  \label{tab:desktopdetail}
  \begin{tabular}{p{4cm}cccccccccccccc}
    \toprule
   \small{\textbf{Indicator}} & \multicolumn{2}{c}{\small{\textbf{$\diameter$ Correlation}}} & \multicolumn{2}{c}{\small{\textbf{\# Significant}}} & \multicolumn{2}{c}{\small{\textbf{Corr.~< .2}}} & \multicolumn{2}{c}{\small{\textbf{Corr.~< .4}}} & \multicolumn{2}{c}{\small{\textbf{Corr.~< .6}}} & \multicolumn{2}{c}{\small{\textbf{Corr.~< .8}}} & \multicolumn{2}{c}{\small{\textbf{Corr.~>= .8}}} \\
   & \tiny{Overview} & \tiny{Detail} & \tiny{Overview} & \tiny{Detail} & \tiny{Overview} & \tiny{Detail} & \tiny{Overview} &\tiny{Detail} & \tiny{Overview} & \tiny{Detail} & \tiny{Overview} & \tiny{Detail} & \tiny{Overview} & \tiny{Detail} \\
    \midrule
    Visibility Count & .65 & .70 & 13 & 35 & 0 & 0 & 0 & 0 & 3 & 10 & 8 & 12 & 2 & 13  \\
    Visibility Seconds & .60 & .69 & 7 & 34 & 0 & 0 & 0 & 0 & 3 & 12 & 3 & 13 & 1 & 9  \\
    Random Movement & .59 & .68 & 17 & 34 & 0 & 0 & 0 & 0& 9 & 12 & 8 & 13 & 0 & 9  \\
    Move in Fragment & .60 & .68 & 26 & 38 & 0 & 0 & 1 & 0 & 13 & 12& 10 & 19 & 2 & 7  \\
    Mouse Over Fragment Count & .58 & .71 & 15 & 34 & 0 & 0 & 0 & 0 & 9 & 8 & 6 & 15 & 0 & 11 \\
    Mouse Over Fragment Seconds & .61 & 66  & 19 & 28 & 0 & 0 & 0 & 0 & 10 & 10 & 6 & 12 & 3 & 6  \\
    Horizontal Movement & .54 & .63 & 11 & 19& 0 & 0 & 0 & 0 & 9 & 9 & 2 & 8 & 0 & 2 \\
    Vertical Movement & .50 & .65 & 4 & 16 & 0 & 0 & 0 & 0 & 3 & 6 & 1 & 9 & 0 & 1  \\ 
    Mouse on Same Y Count & .55 & .70 & 12 & 40 & 0 & 0 & 2 & 0 & 6 & 10 & 4 & 17 & 0 & 13  \\
    Mouse on Same Y Seconds & .59 & .69 & 18 & 38 & 0 & 0 & 0 & 0 & 14 & 15 & 2 & 15 & 2 & 8 \\
    Contact in Fragment & .73 & .64 & 17 & 15 & 0 & 0 & 0 & 0 & 4 & 7 & 6 & 5 & 7 & 3 \\
  \bottomrule
\end{tabular}
\end{table*}

\subsection{Results}
\label{sec:results}
In this section we analyze the results of our study, split up into insights for desktop (cf. Section \ref{sec:resultsdesktop}) and mobile (cf. Section \ref{sec:resultsmobile}) environments. The results are based on a correlation analysis using Pearson's $r$\footnote{\revi{Our data complies with the prerequisites for computation of Pearson correlation.}} 
between the implicit interest indicators and the users' explicit interest indication per article provided in the questionnaire. 
\rev{Please note that due to the moderate number of participants (especially as split up into desktop and mobile users), the identified correlations are considered as first insights to show an overall tendency. In order to gain a stable understanding, further analyses are necessary (see Section \ref{sec:impact})}.
Correlations \rev{were first analyzed} per article per indicator per user and then summarized to identify indicators that are specifically relevant for a larger number of users. However, we argue that the significance of certain indicators and their capability to predict user interest is highly individual and might thus differ from user to user. Thus, we utilize our results to \rev{suggest}
a \rev{first} generalizable formula which can then be individually weighed in order to form individual user models (cf. Section \ref{sec:usermodel}).

\subsubsection{Reading Behavior on Desktop Devices}
\label{sec:resultsdesktop}
For the desktop setting, we analyzed 14 indicators in total whereby \textit{Select}, \textit{Cut \& Copy} and \textit{Zoom} had no occurrences for any of the 74 desktop users and were thus excluded. The results of the correlation analysis for the remaining 11 indicators, summarized for all desktop users, distinguishing between the news overview page and the article detail pages are shown in Table \ref{tab:resultsoverviewdesktop}. 
This distinction is necessary because some indicators are arguably more meaningful for analysis of reading behavior on the overview or detail page (e.g., \textit{Mouse on Same Y} is probably more conclusive for the overview page where there are several different fragments).
We report Pearson's $r$ as well as the related statistical significance ($p$). Statistically significant correlations are marked with * (.05 level) or ** (.001 level). As Table \ref{tab:resultsoverviewdesktop} shows, we mostly found (at least slightly) positive correlations. 
The correlation values themselves  seem relatively low (even for the correlations that were found to be statistically significant). This however is only the case for the average over all users. The analysis on the level of the individual user reports a different view.

\begin{table}[!b]
  \caption{Correlation (Pearson's \textit{r}) overview for all indicators in the mobile setting. * or ** denote statistical significance (on the .05 or .001 level).}
  \label{tab:resultsoverviewmobile}
  \begin{tabular}{p{3.2cm}p{0.92cm}p{0.92cm}p{0.92cm}p{0.92cm}}
    \toprule
    \textbf{Indicator} & \textbf{r}\newline \tiny{Overview Page} & \textbf{p} \newline \tiny{Overview Page} &  \textbf{r} \newline \tiny{Detail Page} & \textbf{p} \newline \tiny{Detail Page}\\
    \midrule
    \textbf{Visibility} & & & \\
    Count & .204 & < .05* & .431 & < .001**\\
    Second & .115 & .073 & .387 & < .001**\\
    \midrule
    \textbf{Contact in Fragment} &  .298 & < .001** & .304 & < .001**\\
    \midrule
    \textbf{Contact on \newline Same Y Count} & .225 & < .05* & .304 & < .001**\\
      \midrule
     \textbf{Visible Before Swipe} & .115 & .073 & .247 & < .001**\\
    \midrule
    \textbf{Visible After Swipe} & -.098 & .108 & .247 & < .001**\\
    \midrule
    \textbf{Skipped While Swipe} & .059 & .227 & . & .\\
    \midrule
    \textbf{Orientation Change Landscape} & . & . & .155 & < .05*\\
    \midrule
    \textbf{Orientation Change Portrait} & . & . & .155 & < .05*\\
  \bottomrule
\end{tabular}
\end{table}

\begin{table*}[!t]
  \caption{Statistically significant correlations of interest indicators with explicit user interest assessment. All values except mean correlation (values between -1 and 1) are in absolute numbers (out of 22 mobile users), measured for the news overview page and detail page. Correlations not statistically significant for overview and detail pages are not included (if significant only for overview or detail page, the values for the respective other are replaced by ``-''). 
  }
  \label{tab:mobiledetail}
  \begin{tabular}{p{4cm}cccccccccccccc}
    \toprule
   \small{\textbf{Indicator}} & \multicolumn{2}{c}{\small{\textbf{$\diameter$ Correlation}}} & \multicolumn{2}{c}{\small{\textbf{\# Significant}}} & \multicolumn{2}{c}{\small{\textbf{Corr.~< .2}}} & \multicolumn{2}{c}{\small{\textbf{Corr.~< .4}}} & \multicolumn{2}{c}{\small{\textbf{Corr.~< .6}}} & \multicolumn{2}{c}{\small{\textbf{Corr.~< .8}}} & \multicolumn{2}{c}{\small{\textbf{Corr.~>= .8}}} \\
   & \tiny{Overview} & \tiny{Detail} & \tiny{Overview} & \tiny{Detail} & \tiny{Overview} & \tiny{Detail} & \tiny{Overview} &\tiny{Detail} & \tiny{Overview} & \tiny{Detail} & \tiny{Overview} & \tiny{Detail} & \tiny{Overview} & \tiny{Detail} \\
    \midrule
    Visibility Count & .71 & - & 5 & - & 0 & - & 0 & - & 2 & - & 1 & - & 2 & -  \\
    Visibility Seconds & .55 & - & 2 & - & 0 & - & 0 & - & 1 & - & 1 & - & 0 & -  \\
    Contact in Fragment & .84 & .73 & 4 & 5 & 0 & 0 & 0 & 0 & 0 & 2 & 2 & 1 & 2 & 2  \\
    Contact on Same Y Count & .65 & - & 2 & - & 0 & - & 0 & - & 1 & - & 1 & - & 0 & -  \\
    Visible Before Swipe & .76 & .64 & 3 & 6 & 0 & 0 & 0 & 0 & 0 & 1 & 2 & 2 & 1 & 3  \\
    Visible After Swipe & .53 & .45 & 1 & 8 & 0 & 0 & 0 & 0 & 1 & 0 & 0 & 6 & 0 & 2  \\
    Skipped While Swipe & -.39 & .64 & 1 & 6 & 1 & 0 & 0 & 0 & 0 & 1 & 0 & 2 & 0 & 3  \\
    Orientation Change Landscape & - & .45 & - & 8 & - & 0 & - & 0 & - & 1 & - & 5 & - & 2  \\
  \bottomrule
  
\end{tabular}
\end{table*}

Thus, we next analyzed the results on the level of the individual (see Table \ref{tab:desktopdetail}). The table reports only the results for users where we found significant correlations between the respective implicit indicator and the related explicit answer the user gave in the questionnaire. For instance, Table \ref{tab:desktopdetail} shows that for the overview page, the indicator \textit{Contact in Fragment} was relevant for 17 of the 74 desktop users. For these 17 users, the mean correlation was .73 (thus, a relatively strong positive correlation). Further, the table shows that for 7 of these 17 users, the correlation was \revi{above} or equal to .8, for 6 users it was between .6 and .8 and for 4 users it was between .4 and .6. As the table shows, the indicators relevant for most users were 
\textit{Move in Fragment} (significant correlation for 26 of 74 users), \textit{Mouse Over Fragment Seconds} (significant correlation for 19 users), \textit{Mouse on Same Y Seconds} (18 users) and \textit{Contact in Fragment} and \textit{Random Movement} (both 17 users).

For the news detail pages more indicators were significant for a larger number of users, e.g., \textit{Mouse on Same Y Count} was significant for more than half of the users (40 of 74), \textit{Move in Fragment} as well as \textit{Mouse on Same Y Seconds} were significant for 38 of 74 users. For the detail pages, all correlations were equal to or above .4.

\subsubsection{Reading Behavior on Mobile Devices}
\label{sec:resultsmobile}
From the 12 mobile indicators in total we did not find any occurrences of \textit{Select}, \textit{Cut \& Copy} and \textit{Zoom} for any of the 22 mobile users. Thus, Table \ref{tab:resultsoverviewmobile} shows the remaining 9 indicators for the mobile setting, again split up into correlations (and related statistical significance) for the overview and the detail pages. With one exception (indicator \textit{Visible After Swipe}), we only found positive correlations. 
For two indicators, there were only occurrences for the news detail pages (\textit{Orientation Change Landscape} and \textit{Orientation Change Portrait}). 

Similar to the desktop setting, we also analyzed the results on the level of the individual user, cf.~Table~\ref{tab:mobiledetail}. This table shows the results only for users where we found significant correlations between the implicit indicators and the users' explicit statements. For instance, the indicator \textit{Visibility} time in seconds was relevant for 5 of 22 users for the overview page. For 2 of these 5 users, we found a correlation greater than or equal to .8, for the remaining 3 the correlation was between .6 and .8. 
For \textit{Orientation Change Landscape} we did not find any significant correlations on the overview page.

For the news detail pages, we did not find any significant correlations for some of the other indicators: \textit{Visibility Count}, \textit{Visibility Seconds}, 
\textit{Contact on Same Y} (although there were occurrences of the related events). For instance, we found significant correlations for 8 of the 22 mobile users between the indicators \textit{Orientation Change Landscape} and \textit{Visible After Swipe} and the related explicit interest statements. Again, the correlations for the users where correlations between indicators and explicit statements were statistically significant, were all at least .4 or higher for the detail pages.

%% file: usermodels.tex
\section{Derived User Models} 
\label{sec:usermodel}
\rev{Based on our first insights} related to implicit interest indicators and their relevance for individual users and groups of users, we \rev{suggest} a generalizable formula as base user model, considering all indicators for which a significant correlation between implicit and explicit interest indicators and statements was found. \rev{Second, we show an exemplary application of this idea by reporting }the computed results for individual instances of the formula for 
certain users (we demonstrate this step for a small sample of representative desktop and mobile users). The formula per user considers only those indicators where significant positive correlations have been found for this particular user. \revi{It can be generally described as a weighted average as it is has been traditionally used for computations of rating predictions in neighborhood-based recommender systems (see e.g., \cite{ning15}). It is explained in further detail as follows}.
Equation \ref{eq:general} shows the computation of predicted interest for a specific user in a certain article where $n$ is the number of significant correlations between implicit indicators and explicit interest statement. Both detail and overview pages are considered for the computation. $v_i$ is the user's normalized value for the implicit indicator. The values are normalized to a value between 0 and 1, based on the user's concrete values for this indicator for all detail or overview pages, see Equation \ref{eq:value}. $corr_i$ is the user's correlation (\rev{in our example} Pearson's $r$) for a specific significant indicator. \rev{We consider this formula a general suggestion that does not necessarily need to be directly related to the indicators used in our study or Pearson's $r$ for correlation. It could easily be adapted to other scenarios.}
\begin{equation}\label{eq:general}
    Pred_{interest} = \frac{\sum_{i=1}^{n}{v_i*corr_i}}{\sum_{i=1}^n{corr_i}}
\end{equation}

\begin{equation}
    \label{eq:value}
    v_i = \frac{v_{orig}-v_{min}}{v_{max}-v_{min}}
\end{equation}

\begin{table}[!b]
\begin{threeparttable}
  \caption{Examples for predicted interest values in comparison with users' explicit statements. We show exemplary predictions for two desktop and two mobile users (one with many, one with only few significant correlations) and two exemplary articles.}
  \label{tab:usermodels}
  \begin{tabular}{p{3.5cm}c|cc|c}
    \toprule 
    \textbf{User} & \multicolumn{2}{c}{\textbf{Article A}} & \multicolumn{2}{c}{\textbf{Article B}} \\
      \midrule
    & \tiny{Predicted} & \tiny{Explicit} & \tiny{Predicted} & \tiny{Explicit}\\   
    \midrule
    \textbf{Desktop} & & & &\\
    U1 - Many sig.~correlations & 0.3 & 0.33 & 0.61 & 0.67\\
    U2 - Few sig.~correlations & 0.02 & 0.33\tnote{1} & 1.00 &  0.83\\ 
    \midrule
    \textbf{Mobile} & & & & \\
    U3 - Many sig.~correlations & 0.01 & 0.17 & 1.00 & 1.00\\
    U4 - Few sig.~correlations & 0.10 & 0.17 & 0.59 & 0.83 \\
  \bottomrule
\end{tabular}
\begin{tablenotes}
\item[1] The value of $0.33$ corresponds to the value $3$ provided by the user which was this user's personal minimum.
\end{tablenotes}
\end{threeparttable}
\end{table}

\rev{To get a feeling for the applicability of this idea, we computed predicted interest values for the articles for the participants of our study. For these samples, we} found a relatively close match between predicted and explicit values in most cases. \rev{This indicates that our approach bears potential to be able to actually predict interest.} \rev{However, further robustness tests need to be conducted to verify the correlations.}
Table \ref{tab:usermodels} shows example results for the predicted interest based on the implicit interest indicators and the related explicitly provided user interest for a small sample of selected users and articles. Thus, the table allows for a comparison between predicted interest based on implicit interest indicators and explicit statements. For reasons of better comparability, we translated the users' statements (provided using a seven-point Likert scale) to a value range of 0 to 1. The value range does not consider a user's individual rating behavior (thus, the range is equal for all users). Table \ref{tab:usermodels} includes an example where this might be misleading at first glance because this user's personal minimum was 3 (instead of 1 as for most of the other users).
The examples displayed in Table \ref{tab:usermodels} were selected as follows: for both environments, desktop and mobile, we selected representative users with either a lot of indicators where we could find significant correlations 
or only very few significant correlations.  
From the predicted interest in certain articles, we later want to be able to derive more common interest in certain topics that are associated with the articles. Further, we envision using the predicted interest in certain articles from reading behavior to derive interest in articles that can be considered similar (either based on meta data or content). For more considerations on future work, consider \revi{Sections \ref{sec:impact}} and \ref{sec:conclusions}.

%% file: impact.tex
\section{Discussion}
\label{sec:impact}
Our \rev{analysis} (and prototype) contributes to the existing body of knowledge on implicit interest indicators as basis for web content recommendation. \revi{It constitutes an update to existing findings of earlier, similar studies for two reasons. First, it revisits known concepts using \emph{up-to-date web technology} (including state of the art frameworks) for the analysis of client-side interactions which also led to a \emph{higher number of indicators}. Second, it \emph{introduces mobile interactions} which have not been considered yet in any of the earlier studies discussed in Section \ref{sec:relatedwork}}. 
\revi{Comparing the mobile setting to the desktop setting, it has to be noted that the mobile setting poses certain challenges. On the one hand it is challenging to precisely and reliable cover the indicators on the huge amount of different mobile devices. On the other hand, fewer indicators can be gained in the mobile setting since no movement-based indicators may be obtained, aggravating a trustworthy derivation of user interests. However, the obtained indicators generally showed a higher correlation and significance in our study setting than the ones on the desktop.}

\rev{While we acknowledge that the reliability of the results of our study might be limited due to the moderate number of participants in the two groups, we consider them important first insights that point to interesting directions and seem to indicate that our approach has potential to be able to predict user interest.}
\rev{For instance, }our \rev{first} results suggest that the \textit{predicted user interest} based on implicit indicators is relatively \textit{close to the users' explicit interest statements}. Thus, the derived user models seem appropriate for further consideration in web content recommender systems. 
Further potential limitations are discussed as follows. 


One issue related to the design of our news page is a potential \textit{bias} based on the \textit{position} of the respective fragment. This bias is reported by several studies (see e.g., \cite{Joachims2005, Craswell2008}) and implies that users are more likely to read text that is positioned at top of a page. A potential solution would have been to randomize the order of the fragments on the overview page which would however have lead to implementation-related difficulties (e.g., for the identification of items skipped while swiping). Further, we (intentionally) \textit{reduced} the content of our news page to \textit{plain text} in order to avoid side effects related to design, position or type of multimedia elements. It is conceivable that for other types of content the significance of individual interest indicators differs from what we report for text content. 

On the news detail pages, our study considered the full text as \textit{one page fragment}. To allow for a more fine-grained analysis, the page could be further split up into a higher number of fragments (this is generally already possible with our prototype and considered for future work). Further, our study investigated a \textit{rather static page}. Contemporary websites are often more dynamic (e.g., single page applications with heavy use of asynchronous data loads). Full support for such websites would require changes in the implementation which might affect the functionality of certain indicators. \revi{Additionally, currently we focused on the individual analysis of certain indicators only, but did not consider certain sequences or combinations of indicators that might exhibit specific semantics. For example, on mobile phones it is often hard to reliably distinguish between clicking and swiping resulting in unintentional clicks that are immediately followed by a click on the back button. Detecting such sequences might help to avoid counting false positives.}


\revi{Finally, a more general limitation is related to the well-known cold start problem (see e.g., \cite{ji15}). Our suggested procedure for the prediction of user interest (see Section \ref{sec:usermodel}) relies on users' explicit feedback to identify their individual implicit indicators on at least one occasion during system use. This is not generally uncommon to tackle the cold start problem. For instance, popular recommendation services such as Netflix\footnote{https://www.netflix.com} rely on explicit feedback which is usually collected upon registration at the system to provide high-quality  recommendations before enough implicit information is available. Nevertheless, we believe that future work should also investigate alternatives. For instance, it might be possible to create a ``default user model'' or ``group model'' comprising only the indicators which have been proven to be relevant for the majority of all users so far. This default model could then be stepwise refined for individual users as more information is revealed to the system during longer-term use. Such a default model would however have to be validated by a study with a high number of users first, in order to make sure the findings can be considered reliable.}

%% file: conclusions.tex
\section{Conclusions}
\label{sec:conclusions}
In this paper, we discussed the \rev{analysis} of implicit indicators for user interest on a web site based on client-side analysis of reading behavior. We presented a prototypical technical infrastructure and concrete application as well as the design and results of an online study with desktop and mobile users. Although client-side analysis of reading behavior and derivation of implicit interest indicators have been researched before, 
our work exceeds the state of the art in several ways \revi{(see Section \ref{sec::introduction})}. 
Our work can be seen as a first \rev{step} in the development of web content recommender systems based on such indicators. 
\rev{Future work should include a subsequent analysis with a higher number of samples in order to gain statistically more reliable results. Additionally, we plan using the approach, indicators and general model structure presented in this paper to further analyze the data with different correlation measures (e.g., Spearman's rank correlation).}
In the future it is \rev{also} envisioned to utilize the user models we \rev{suggested} in combination with extraction of information from the content itself. Our approach \rev{potentially} enables a system to determine which content is of interest for a user. Through further information extraction, a system could derive relevant meta-data (e.g., length or source of a text or what genre it belongs to) and then recommend similar content. 
We believe that implicit, unobtrusive client-side interest analysis has high potential to provide high-quality recommendations \revi{on intelligent web sites} while not requiring too much activity and attention from the user. 




%% file: umap.bbl

\begin{thebibliography}{27}


\ifx \showCODEN    \undefined \def \showCODEN     #1{\unskip}     \fi
\ifx \showDOI      \undefined \def \showDOI       #1{#1}\fi
\ifx \showISBNx    \undefined \def \showISBNx     #1{\unskip}     \fi
\ifx \showISBNxiii \undefined \def \showISBNxiii  #1{\unskip}     \fi
\ifx \showISSN     \undefined \def \showISSN      #1{\unskip}     \fi
\ifx \showLCCN     \undefined \def \showLCCN      #1{\unskip}     \fi
\ifx \shownote     \undefined \def \shownote      #1{#1}          \fi
\ifx \showarticletitle \undefined \def \showarticletitle #1{#1}   \fi
\ifx \showURL      \undefined \def \showURL       {\relax}        \fi
\providecommand\bibfield[2]{#2}
\providecommand\bibinfo[2]{#2}
\providecommand\natexlab[1]{#1}
\providecommand\showeprint[2][]{arXiv:#2}

\bibitem[\protect\citeauthoryear{Bilenko and Richardson}{Bilenko and
  Richardson}{2011}]%
        {Bilenko2011}
\bibfield{author}{\bibinfo{person}{Mikhail Bilenko} {and}
  \bibinfo{person}{Matthew Richardson}.} \bibinfo{year}{2011}\natexlab{}.
\newblock \showarticletitle{Predictive Client-side Profiles for Personalized
  Advertising}. In \bibinfo{booktitle}{\emph{Proceedings of the 17th ACM SIGKDD
  International Conference on Knowledge Discovery and Data Mining}}
  \emph{(\bibinfo{series}{KDD '11})}. \bibinfo{publisher}{ACM},
  \bibinfo{address}{New York, NY, USA}, \bibinfo{pages}{413--421}.
\newblock
\showISBNx{978-1-4503-0813-7}
\urldef\tempurl%
\url{https://doi.org/10.1145/2020408.2020475}
\showDOI{\tempurl}


\bibitem[\protect\citeauthoryear{Claypool, Le, Wased, and Brown}{Claypool
  et~al\mbox{.}}{2001}]%
        {Claypool2001}
\bibfield{author}{\bibinfo{person}{Mark Claypool}, \bibinfo{person}{Phong Le},
  \bibinfo{person}{Makoto Wased}, {and} \bibinfo{person}{David Brown}.}
  \bibinfo{year}{2001}\natexlab{}.
\newblock \showarticletitle{Implicit Interest Indicators}. In
  \bibinfo{booktitle}{\emph{Proceedings of the 6th International Conference on
  Intelligent User Interfaces}} \emph{(\bibinfo{series}{IUI '01})}.
  \bibinfo{publisher}{ACM}, \bibinfo{address}{New York, NY, USA},
  \bibinfo{pages}{33--40}.
\newblock
\showISBNx{1-58113-325-1}


\bibitem[\protect\citeauthoryear{Craswell, Zoeter, Taylor, and Ramsey}{Craswell
  et~al\mbox{.}}{2008}]%
        {Craswell2008}
\bibfield{author}{\bibinfo{person}{Nick Craswell}, \bibinfo{person}{Onno
  Zoeter}, \bibinfo{person}{Michael Taylor}, {and} \bibinfo{person}{Bill
  Ramsey}.} \bibinfo{year}{2008}\natexlab{}.
\newblock \showarticletitle{An Experimental Comparison of Click Position-Bias
  Models}. In \bibinfo{booktitle}{\emph{Proceedings of the 2008 International
  Conference on Web Search and Data Mining}}.
\newblock


\bibitem[\protect\citeauthoryear{Faucher, McLoughlin, and Wunschel}{Faucher
  et~al\mbox{.}}{2011}]%
        {Faucher2011}
\bibfield{author}{\bibinfo{person}{Joshua Faucher}, \bibinfo{person}{Brendan
  McLoughlin}, {and} \bibinfo{person}{Jennifer Wunschel}.}
  \bibinfo{year}{2011}\natexlab{}.
\newblock \showarticletitle{Implicit web user interest}.
\newblock \bibinfo{journal}{\emph{Technical Report MQP-CEW-1101, Worcester
  Polytechnic Institute}} (\bibinfo{year}{2011}).
\newblock


\bibitem[\protect\citeauthoryear{Flesch}{Flesch}{1948}]%
        {flesch48}
\bibfield{author}{\bibinfo{person}{Rudolph Flesch}.}
  \bibinfo{year}{1948}\natexlab{}.
\newblock \showarticletitle{A New Readability Yardstick}.
\newblock \bibinfo{journal}{\emph{Journal of Applied Psychology}}
  \bibinfo{volume}{32}, \bibinfo{number}{3} (\bibinfo{year}{1948}),
  \bibinfo{pages}{221--233}.
\newblock


\bibitem[\protect\citeauthoryear{Goecks and Shavlik}{Goecks and
  Shavlik}{2000}]%
        {Goecks2000}
\bibfield{author}{\bibinfo{person}{Jeremy Goecks} {and} \bibinfo{person}{Jude
  Shavlik}.} \bibinfo{year}{2000}\natexlab{}.
\newblock \showarticletitle{Learning Users' Interests by Unobtrusively
  Observing Their Normal Behavior}. In \bibinfo{booktitle}{\emph{Proceedings of
  the 5th International Conference on Intelligent User Interfaces}}
  \emph{(\bibinfo{series}{IUI '00})}. \bibinfo{publisher}{ACM},
  \bibinfo{address}{New York, NY, USA}, \bibinfo{pages}{129--132}.
\newblock
\showISBNx{1-58113-134-8}
\urldef\tempurl%
\url{https://doi.org/10.1145/325737.325806}
\showDOI{\tempurl}


\bibitem[\protect\citeauthoryear{Hauger}{Hauger}{2008}]%
        {Hauger2008}
\bibfield{author}{\bibinfo{person}{David Hauger}.}
  \bibinfo{year}{2008}\natexlab{}.
\newblock \showarticletitle{Fine-Grained User Models by Means of Asynchronous
  Web Technologies}. In \bibinfo{booktitle}{\emph{16th Workshop on Adaptivity
  and User Modeling in Interactive Systems}}.
\newblock


\bibitem[\protect\citeauthoryear{Hauger}{Hauger}{2009}]%
        {Hauger2009b}
\bibfield{author}{\bibinfo{person}{David Hauger}.}
  \bibinfo{year}{2009}\natexlab{}.
\newblock \showarticletitle{Using Asynchronous Client-Side User Monitoring to
  Enhance User Modeling in Adaptive E-Learning Systems}. In
  \bibinfo{booktitle}{\emph{Proceedings of the Workshop on Adaptation and
  Personalization for Web 2.0}}.
\newblock


\bibitem[\protect\citeauthoryear{Hauger, Paramythis, and Weibelzahl}{Hauger
  et~al\mbox{.}}{2011}]%
        {Hauger2011}
\bibfield{author}{\bibinfo{person}{David Hauger}, \bibinfo{person}{Alexandros
  Paramythis}, {and} \bibinfo{person}{Stephan Weibelzahl}.}
  \bibinfo{year}{2011}\natexlab{}.
\newblock \showarticletitle{Using browser interaction data to determine page
  reading behavior}. In \bibinfo{booktitle}{\emph{International Conference on
  User Modeling, Adaptation, and Personalization}}. \bibinfo{pages}{147--158}.
\newblock


\bibitem[\protect\citeauthoryear{Hauger and Van~Velsen}{Hauger and
  Van~Velsen}{2009}]%
        {Hauger2009}
\bibfield{author}{\bibinfo{person}{David Hauger} {and} \bibinfo{person}{Lex
  Van~Velsen}.} \bibinfo{year}{2009}\natexlab{}.
\newblock \showarticletitle{Analyzing Client-Side Interactions to Determine
  Reading Behavior}. In \bibinfo{booktitle}{\emph{ABIS 2009}}.
  \bibinfo{address}{Darmstadt, Germany}, \bibinfo{pages}{11--16}.
\newblock


\bibitem[\protect\citeauthoryear{Hijikata}{Hijikata}{2004}]%
        {Hijikata2004}
\bibfield{author}{\bibinfo{person}{Yoshinori Hijikata}.}
  \bibinfo{year}{2004}\natexlab{}.
\newblock \showarticletitle{Implicit User Profiling for on Demand Relevance
  Feedback}. In \bibinfo{booktitle}{\emph{Proceedings of the 9th International
  Conference on Intelligent User Interfaces}} \emph{(\bibinfo{series}{IUI
  '04})}. \bibinfo{publisher}{ACM}, \bibinfo{address}{New York, NY, USA},
  \bibinfo{pages}{198--205}.
\newblock
\showISBNx{1-58113-815-6}
\urldef\tempurl%
\url{https://doi.org/10.1145/964442.964480}
\showDOI{\tempurl}


\bibitem[\protect\citeauthoryear{Jawaheer, Szomszor, and Kostkova}{Jawaheer
  et~al\mbox{.}}{2010}]%
        {jawaheer2010}
\bibfield{author}{\bibinfo{person}{Gawesh Jawaheer}, \bibinfo{person}{Martin
  Szomszor}, {and} \bibinfo{person}{Patty Kostkova}.}
  \bibinfo{year}{2010}\natexlab{}.
\newblock \showarticletitle{Comparison of implicit and explicit feedback from
  an online music recommendation service}. In
  \bibinfo{booktitle}{\emph{{Proceedings of the 1st International Workshop on
  information heterogeneity and fusion in recommender systems}}}. ACM,
  \bibinfo{pages}{47--51}.
\newblock


\bibitem[\protect\citeauthoryear{Ji and Shen}{Ji and Shen}{2015}]%
        {ji15}
\bibfield{author}{\bibinfo{person}{Ke Ji} {and} \bibinfo{person}{Hong Shen}.}
  \bibinfo{year}{2015}\natexlab{}.
\newblock \showarticletitle{Addressing Cold-Start: Scalable Recommendation with
  Tags and Keywords}.
\newblock \bibinfo{journal}{\emph{Knowledge-Based Systems}}
  \bibinfo{volume}{83} (\bibinfo{year}{2015}), \bibinfo{pages}{42--50}.
\newblock


\bibitem[\protect\citeauthoryear{Joachims, Granka, Pan, Hembrooke, and
  Gay}{Joachims et~al\mbox{.}}{2005}]%
        {Joachims2005}
\bibfield{author}{\bibinfo{person}{Thorsten Joachims}, \bibinfo{person}{Laura
  Granka}, \bibinfo{person}{Bing Pan}, \bibinfo{person}{Helen Hembrooke}, {and}
  \bibinfo{person}{Geri Gay}.} \bibinfo{year}{2005}\natexlab{}.
\newblock \showarticletitle{Accurately Interpreting Clickthrough Data as
  Implicit Feedback}. In \bibinfo{booktitle}{\emph{Proceedings of the 28th
  Annual International ACM SIGIR Conference on Research and Development in
  Information Retrieval}}. \bibinfo{pages}{154--161}.
\newblock


\bibitem[\protect\citeauthoryear{Kellar, Watters, Duffy, and Shepherd}{Kellar
  et~al\mbox{.}}{2004}]%
        {Kellar2004}
\bibfield{author}{\bibinfo{person}{Melanie Kellar}, \bibinfo{person}{Carolyn
  Watters}, \bibinfo{person}{Jack Duffy}, {and} \bibinfo{person}{Michael
  Shepherd}.} \bibinfo{year}{2004}\natexlab{}.
\newblock \showarticletitle{Effect of task on time spent reading as an implicit
  measure of interest}.
\newblock \bibinfo{journal}{\emph{Proceedings of the American Society for
  Information Science and Technology}} \bibinfo{volume}{41},
  \bibinfo{number}{1} (\bibinfo{year}{2004}), \bibinfo{pages}{168--175}.
\newblock


\bibitem[\protect\citeauthoryear{K{\v{r}}{\'\i}{\v{z}}}{K{\v{r}}{\'\i}{\v{z}}}{2012}]%
        {Kvrivz2012}
\bibfield{author}{\bibinfo{person}{Jakub K{\v{r}}{\'\i}{\v{z}}}.}
  \bibinfo{year}{2012}\natexlab{}.
\newblock \showarticletitle{Keyword Extraction Based on Implicit Feedback.}
\newblock \bibinfo{journal}{\emph{Information Sciences \& Technologies:
  Bulletin of the ACM Slovakia}} \bibinfo{volume}{4}, \bibinfo{number}{2}
  (\bibinfo{year}{2012}).
\newblock


\bibitem[\protect\citeauthoryear{Mueller and Lockerd}{Mueller and
  Lockerd}{2001}]%
        {Mueller2001}
\bibfield{author}{\bibinfo{person}{Florian Mueller} {and}
  \bibinfo{person}{Andrea Lockerd}.} \bibinfo{year}{2001}\natexlab{}.
\newblock \showarticletitle{Cheese: tracking mouse movement activity on
  websites, a tool for user modeling}. In \bibinfo{booktitle}{\emph{CHI'01
  extended abstracts on Human factors in computing systems}}. ACM,
  \bibinfo{pages}{279--280}.
\newblock


\bibitem[\protect\citeauthoryear{Ning, Desrosiers, and Karypis}{Ning
  et~al\mbox{.}}{2015}]%
        {ning15}
\bibfield{author}{\bibinfo{person}{Xia Ning}, \bibinfo{person}{Christian
  Desrosiers}, {and} \bibinfo{person}{George Karypis}.}
  \bibinfo{year}{2015}\natexlab{}.
\newblock \showarticletitle{A Comprehensive Survey of Neighborhood-Based
  Recommendation Methods}.
\newblock In \bibinfo{booktitle}{\emph{Recommender Systems Handbook}
  (\bibinfo{edition}{2nd} ed.)}, \bibfield{editor}{\bibinfo{person}{Francesco
  Ricci}, \bibinfo{person}{Lior Rokach}, {and} \bibinfo{person}{Bracha
  Shapira}} (Eds.).
\newblock


\bibitem[\protect\citeauthoryear{Rastegari and Shamsuddin}{Rastegari and
  Shamsuddin}{2010}]%
        {Rastegari2010}
\bibfield{author}{\bibinfo{person}{Hamid Rastegari} {and}
  \bibinfo{person}{Siti~Mariyam Shamsuddin}.} \bibinfo{year}{2010}\natexlab{}.
\newblock \showarticletitle{Web search personalization based on browsing
  history by artificial immune system}.
\newblock \bibinfo{journal}{\emph{International Journal of Advances in Soft
  Computing and Its Applications}} \bibinfo{volume}{2}, \bibinfo{number}{3}
  (\bibinfo{year}{2010}).
\newblock


\bibitem[\protect\citeauthoryear{Shen, Tan, and Zhai}{Shen
  et~al\mbox{.}}{2005}]%
        {Shen2005}
\bibfield{author}{\bibinfo{person}{Xuehua Shen}, \bibinfo{person}{Bin Tan},
  {and} \bibinfo{person}{ChengXiang Zhai}.} \bibinfo{year}{2005}\natexlab{}.
\newblock \showarticletitle{Context-sensitive information retrieval using
  implicit feedback}. In \bibinfo{booktitle}{\emph{Proceedings of the 28th
  annual international ACM SIGIR conference on Research and development in
  information retrieval}}. ACM, \bibinfo{pages}{43--50}.
\newblock


\bibitem[\protect\citeauthoryear{Smucker, Guo, and Toulis}{Smucker
  et~al\mbox{.}}{2014}]%
        {Smucker2014}
\bibfield{author}{\bibinfo{person}{Mark~D Smucker},
  \bibinfo{person}{Xiaoyu~Sunny Guo}, {and} \bibinfo{person}{Andrew Toulis}.}
  \bibinfo{year}{2014}\natexlab{}.
\newblock \showarticletitle{Mouse movement during relevance judging:
  implications for determining user attention}. In
  \bibinfo{booktitle}{\emph{Proceedings of the 37th international ACM SIGIR
  conference on Research \& development in information retrieval}}. ACM,
  \bibinfo{pages}{979--982}.
\newblock


\bibitem[\protect\citeauthoryear{Smucker and Jethani}{Smucker and
  Jethani}{2010}]%
        {smucker2010}
\bibfield{author}{\bibinfo{person}{Mark~D. Smucker} {and}
  \bibinfo{person}{Chandra~P. Jethani}.} \bibinfo{year}{2010}\natexlab{}.
\newblock \showarticletitle{Human Performance and Retrieval Precision
  Revisited}. In \bibinfo{booktitle}{\emph{Proceedings of SIGIR'10}}.
  \bibinfo{address}{Geneva, Switzerland}, \bibinfo{pages}{595--602}.
\newblock


\bibitem[\protect\citeauthoryear{Stephanidis, Paramythis, Karagiannidis, and
  Savidis}{Stephanidis et~al\mbox{.}}{1997}]%
        {Stephanidis1997}
\bibfield{author}{\bibinfo{person}{Constantine Stephanidis},
  \bibinfo{person}{Alexandros Paramythis}, \bibinfo{person}{Charalampos
  Karagiannidis}, {and} \bibinfo{person}{Anthony Savidis}.}
  \bibinfo{year}{1997}\natexlab{}.
\newblock \showarticletitle{Supporting interface adaptation: the AVANTI
  Web-Browser}. In \bibinfo{booktitle}{\emph{3rd ERCIM Workshop on User
  Interfaces for All}}.
\newblock


\bibitem[\protect\citeauthoryear{Torres and Hernando}{Torres and
  Hernando}{2008}]%
        {Torres2008}
\bibfield{author}{\bibinfo{person}{Luis A~Leiva Torres} {and}
  \bibinfo{person}{Roberto~Vivo Hernando}.} \bibinfo{year}{2008}\natexlab{}.
\newblock \showarticletitle{A gesture inference methodology for user evaluation
  based on mouse activity tracking}. In \bibinfo{booktitle}{\emph{IHCI 2008,
  Proceedings of the IADIS International Conference on Interfaces and Human
  Computer Interaction, Amsterdam, The Netherlands}}.
\newblock


\bibitem[\protect\citeauthoryear{Velayathan and Yamada}{Velayathan and
  Yamada}{2007}]%
        {Velayathan2007}
\bibfield{author}{\bibinfo{person}{Ganesan Velayathan} {and}
  \bibinfo{person}{Seiji Yamada}.} \bibinfo{year}{2007}\natexlab{}.
\newblock \showarticletitle{Can we find common rules of browsing behavior}. In
  \bibinfo{booktitle}{\emph{QUERY LOG ANALYSIS, WWW'07: Query Log Analysis:
  Social and Technological Challenges, A workshop at the 16th International
  Conference on World Wide Web}}.
\newblock


\bibitem[\protect\citeauthoryear{Zahoor, Bedekar, and Kosamkar}{Zahoor
  et~al\mbox{.}}{2014}]%
        {Zahoor2014}
\bibfield{author}{\bibinfo{person}{Saniya Zahoor}, \bibinfo{person}{Mangesh
  Bedekar}, {and} \bibinfo{person}{Pranali~K Kosamkar}.}
  \bibinfo{year}{2014}\natexlab{}.
\newblock \showarticletitle{User implicit interest indicators learned from the
  browser on the client side}. In \bibinfo{booktitle}{\emph{Proceedings of the
  2014 International Conference on Information and Communication Technology for
  Competitive Strategies}}. ACM, \bibinfo{pages}{57}.
\newblock


\bibitem[\protect\citeauthoryear{Zahoor, Rajput, Bedekar, and Kosamkar}{Zahoor
  et~al\mbox{.}}{2015}]%
        {Zahoor2015}
\bibfield{author}{\bibinfo{person}{Saniya Zahoor},
  \bibinfo{person}{Digvijaysingh Rajput}, \bibinfo{person}{Mangesh Bedekar},
  {and} \bibinfo{person}{Pranali Kosamkar}.} \bibinfo{year}{2015}\natexlab{}.
\newblock \showarticletitle{Capturing, understanding and interpreting user
  interactions with the browser as implicit interest indicators}. In
  \bibinfo{booktitle}{\emph{Pervasive Computing (ICPC), 2015 International
  Conference on}}. IEEE, \bibinfo{pages}{1--6}.
\newblock


\end{thebibliography}
